# Opening a hydrophobic gate: the nicotinic acetylcholine receptor as an example

**Sarah E. Rogers[1,2], Kaihsu Tai[1], Oliver Beckstein[1,3] and Mark S.P. Sansom[1]**

[1]Department of Biochemistry, University of Oxford, South Parks Road, Oxford OX1 3QU, United Kingdom

[2]Chemical Biology Sub-Department, University of Oxford, Chemistry Research Laboratory, 12 Mansfield Road, Oxford OX1 3TA, United Kingdom

[3]Merton College, Merton Street, Oxford OX1 4JD, United Kingdom
**E-mail:** mark.sansom@bioch.ox.ac.uk

**Running title:** Opening a hydrophobic gate

**Abstract.** To what extent must a hydrophobic gate expand for the channel to count as open? We address this question using the nicotinic acetylcholine receptor (nAChR) as the exemplar. The nAChR is an integral membrane protein which forms a cation selective channel gated by neurotransmitter binding to its extracellular domain. A hydrophobic gating model has been proposed for the nAChR, whereby the pore is incompletely occluded in the closed state channel, with a narrow hydrophobic central gate region which presents an energetic barrier to ion permeation. The nAChR pore is lined by a parallel bundle of five M2 α-helices, with the gate formed by three rings of hydrophobic sidechains (9′, 13′, and 17′ of M2). A number of models have been proposed to describe the nature of the conformational change underlying the closed to open transition of the nAChR. These models involve different degrees of M2 helix displacement, rotation, and/or kinking. In this study, we use a simple pore expansion method (previously used to model opening of potassium channels) to generate a series of progressively wider models of the nAChR transmembrane domain. Continuum electrostatics calculations are used to assess the change in the barrier height of the hydrophobic gate as a function of pore expansion. The results suggest that an increase in radius of $\Delta r \approx 1.5$ Å is sufficient to functionally open the pore without, for example, a requirement for rotation of the M2 helices. This is evaluated in the context of current mutational and structural data on the nAChR and its homologues.

**Keywords:** nicotinic acetylcholine receptor; electrostatics; ion channel; hydrophobic gating; nanopore

**1. Introduction**

Channels and nanopores in membranes are of considerable interest from both a biological (Hille, 2001; Ashcroft, 2000) and a nanotechnological (Fortina *et al.*, 2005; Lu *et al.*, 2006; Archakov and Ivanov, 2007) perspective. In particular, they may provide selective transport across membranes of water, ions, or other solutes (including polymers). Many biological nanopores (e.g. ion channels) are gated, that is they can be switched between functionally open (i.e. ion permeable) and closed (i.e. ion impermeable) states in a controlled fashion. Understanding the underlying molecular design principles of channel gates is of interest if nanopores are to be used in a technological setting (Chen *et al.*, 2008). Consequently, there is considerable interest in understanding the principles of pore gating.

One possible mechanism is that of hydrophobic gating. This proposes that the closed state of the channel may not be completely occluded, but is sufficiently hydrophobic and narrow to energetically exclude the passage of ions and/or water molecules. This concept has been extensively explored in the context of model nanopores (Beckstein *et al.*, 2001; Beckstein and Sansom, 2003, 2004; Beckstein *et al.*, 2004) and certain ion channels (Beckstein *et al.*, 2003; Anishkin and Sukharev, 2004). In particular, it has been suggested to be the mechanistic basis behind gating in the nicotinic acetylcholine receptor (nAChR) and related channels (Amiri *et al.*, 2005; Beckstein and Sansom, 2006).

The nicotinic acetylcholine receptor is an integral membrane protein belonging to the ligand-gated ion channel superfamily of neurotransmitter receptors (Lester, 1992; Lester *et al.*, 2004). Activation occurs on extracellular binding of two acetylcholine molecules, leading to a subsequent transition from a closed to an open state. The nAChR is pentameric, with the five subunits surrounding a central pore (Miyazawa *et al.*, 2003; Unwin, 2005) (figure 1). The immediate lining of the pore is formed by a bundle of five parallel M2 segments, which adopt an α-helical conformation. This transmembrane (TM) pore domain has been extensively probed to elucidate the position of the gate (Lester, 1992; Lester *et al.*, 2004; Corringer *et al.*, 2000) which has been suggested to be formed by the sidechains of the hydrophobic amino acid residues 9′, 13′ and 17′ (figure 2).

Molecular dynamics (MD) and related simulation methods have been applied to understand a number of properties of ion channels (Roux and Schulten, 2004) and of nanopores (Hummer *et al.*, 2001; Allen *et al.*, 2003; Cruz-Chu *et al.*, 2006; Corry, 2008; Bratko *et al.*, 2007; Takaiwa *et al.*, 2008). Using cryoelectron microscopy derived structure of the *Torpedo marmorata* nAChR and models based upon this (Amiri *et al.*, 2005), MD simulations have suggested that conformational changes in the vicinity of the hydrophobic gate may underpin the mechanism of activation (Amiri *et al.*, 2005; Beckstein and Sansom, 2006; Corry, 2006). Both twisting (Taly *et al.*, 2005; Liu *et al.*, 2008) and tilting (Cheng *et al.*, 2007) motions of the M2 helices have been proposed, on the basis of MD and related studies, to underlie channel gating. However, it has also been suggested that a relatively small increase in pore radius may sufficiently reduce the energy barrier to permeation of a sodium ion across the gate region to switch the channel from a functionally closed to a functionally open state. Indeed, recent experimental studies (Cymes and Grosman, 2008) suggest that the increase in nAChR pore radius underlying channel opening may involve only a limited conformational change so as to exploit the steep dependence of ion conduction on pore radius as seen in simulations of model hydrophobic nanopores.

In the current study, we explore the relationship between the pore radius and the electrostatic potential energy profile for sodium ions along the pore axis of models of the nAChR. We use a simple pore expansion method to generate a series of progressively wider models of the nAChR transmembrane domain. Continuum electrostatics calculations are used to assess the change in the barrier height of the hydrophobic gate as a function of pore expansion. The results confirm the hypothesis that a relatively small perturbation of the hydrophobic gate may be sufficient to open the channel. We do not aim to assert that nAChR opens as suggested by our pore expansion method, but that the results here give a flavour of the relationship between pore radius and energy barrier for protein channels in general.

## 2. Methods

### 2.1. Channel structures

Two cryoelectron microscopy structures of the nAChR structures were used: PDB codes 1OED and 2BG9. The transmembrane domains, used in subsequent calculations, were defined as residues 211-437 (α chain), 217-466 (β chain), 225-484 (δ chain), and 219-476 (γ chain). In silico 'mutation' of selected sidechains by substitution with alanine was carried out with PyMOL (http://pymol.sourceforge.net/). Structures were visualized and rendered with VMD and the Tachyon ray tracer (Humphrey et al., 1996) and pore radii were calculated using HOLE (Smart et al., 1996).

*2.2. Poisson–Boltzmann energy calculation*

Electrostatic energy profiles were generated using the Adaptive Poisson–Boltzmann Solver (APBS; http://apbs.sourceforge.net/) program (Baker *et al.*, 2001), in combination with PDB2PQR (Dolinsky *et al.*, 2004) and the AMBER99 forcefield (Wang *et al.*, 2000). APBS was used to calculate the electrostatic field on a grid of points around the transmembrane pore. From preliminary calculations of Poisson–Boltzmann energies through the pore region, a grid of $161 \times 161 \times 193$ points, with spacings of 0.75 Å in *xy* and 0.65 Å in *z* (where z is the long axis of the pore) was found to be suitable for the APBS calculations. The temperature and ambient electrolyte concentration were set to 300 K and 0.15 M respectively, the latter chosen so as to be comparable to physiological ionic concentrations. Having solved the Poisson-Boltzmann calculation on a grid surrounding the transbilayer pore, a sodium ion (assigned a Born radius 1.68 Å (Hummer *et al.*, 1997)) was placed at 0.75 Å intervals along the pore axis as defined by HOLE and the electrostatic free energy of solvation (the 'Born energy') was evaluated at each position.

*2.3. Increasing the pore radius*

The M2 TM helices of the 2BG9 nAChR structure were pushed incrementally outwards using three van der Waal spheres placed at approximately regular 10 Å intervals along the pore (i.e. z) axis. Spheres of increasing size (controlled by the Lennard-Jones length parameter, σ) were created. A sphere with $\sigma_0 = 0.416$ nm did not perturb the pore (for comparison, the σ of a united atom methane group from the 43a1 GROMOS96 forcefield (Scott *et al.*, 1999) is 0.371 nm), so we quote the dimensionless expansion parameter $\rho = \sigma/\sigma_0$ to describe the different spheres used throughout this study. Thus, spheres of increasing radius from $\rho = 1$ to $\rho = 7.2$ were used to progressively increase in the M2 pore radius, each expansion of the sphere radii being followed by energy minimisation. Energy minimisation was carried out using a steepest-descent algorithm, as implemented in GROMACS 3.3.1 (Lindahl *et al.*, 2001) ([www.gromacs.org](www.gromacs.org)). The maximum number of convergence steps was set to 10000. The maximum force parameter, below which energy minimization ceased, was set to $< 1.0$ kJ mol$^{-1}$ nm$^{-1}$. As $\rho$ increases, more and more residues in M2 occupied unfavourable positions in the Ramachandran plot (data not shown). Since the the rest of the protein were held in restraint, the M2 helices still assumed conformations representative of a protein pore (figure 5).

**3. Results**

*3.1. Comparison of closed state structures*

In order to evaluate the open/closed state of the hydrophobic gate in nAChR pore structures and models of the TM region were created from which we measured two properties of the pores: (i) the pore radius profile; and (ii) the electrostatic Poisson-Boltzmann free energy profile. As discussed previously (Beckstein *et al.*, 2004; Amiri *et al.*, 2005) the latter provides a first approximation to the free energy profile of a cation moving along the pore axis.

There are two structures for the presumed closed state of the *Torpedo* nAChR protein, both determined by cryoelectron microscopy, which we will refer to by their PDB codes, namely 1OED (Miyazawa *et al.*, 2003) and 2BG9 (Unwin, 2005). The resolution of both structures is 4 Å, but the 2BG9 structure was refined and so may be viewed as a more accurate model. However, given the resolution of the data, it is valuable to estimate the effect of the changes in (closed state) model structure on the pore radius and electrostatic profiles.

The pore radius profiles for the two structures (1OED and 2BG9) are shown in figure 2, and are broadly similar. However, for the 1OED structure the minimum radius ($r = 2.9$ Å) is in the vicinity of the 13′ and 17′ rings of hydrophobic sidechains (Table 1) at $z = 6.0$ Å, whereas for the 2BG9 pore the minimum radius ($r = 2.4$ Å) was located at $z = 1.3$ Å, corresponding to the 13′ sidechain ring.

In both cases the electrostatic energy profiles (for a $Na^+$ ion) exhibit maxima corresponding to the narrowest regions of the pore. In general, we consider that any barrier $> 2kT$ as significant, for below this threshold an ion could readily overcome the barrier by small thermal fluctuations. The $r = 2.9$ Å 17′ ring of the 1OED pore presents an electrostatic barrier of ~$10kT$ whereas for 2BG9 the $r = 2.4$ Å 13′ sidechain ring generates a barrier of ~ $12kT$. In both cases there is also a barrier in the vicinity of the 2′ and 6′ sidechain rings. However, as this corresponds to small, polar sidechains (e.g. serine and threonine) it is likely that this barrier will disappear due to relaxation of sidechain conformations occurring during passage of an ion (Sansom, 1992; Beckstein and Sansom, 2006).

Overall, comparison of the two structures suggests that in both cases a significant barrier to cation permeation is presented by narrow rings of hydrophobic sidechains, but that the exact height and location of the barrier is sensitive to small differences in the experimentally determined structure.

*3.2. Mutating the pore*

It has been suggested that the nAChR pore region lined with rings of hydrophobic sidechains of residues 9′, 13′, and 17′ may be involved in the gating mechanism of the nAChR (Miyazawa *et al.*, 2003). As discussed above, this region corresponds to the central constriction and electrostatic energy barrier of the pore. To investigate how changes in the pore radius between 9′ to 17′ may alter the electrostatic potential energy profile of the $Na^+$ ion, sidechains in this region were successively removed by *in silico* 'mutation' of selected residues to alanine. Thus, for each sidechain ring, all five sidechains were replaced by that of alanine (a methyl group) resulting in a wider pore in the vicinity of that ring. We note that similar mutations have been used experimentally to probe the structure/function relationships of nAChR channels (e.g. (Labarca *et al.*, 1995)). The resulting radius and electrostatic potential energy profiles are displayed in figure 2. It can be seen that sidechain substitutions to alanine have a major effect on the electrostatic energy profile for residue 13′, in both the 1OED and 2BG9 structures. Substitutions at 9′ and 17′ also had a noticeable effect. Alanine substitutions at other positions did not substantially alter the barrier heights (data not shown).

This analysis suggests that the hydrophobic rings at 9′, 13′, and 17′ did indeed form the energy barriers to sodium ion permeation in the closed state (for both structures). For example, alanine substitution at 13′ in the 2BG9 pore resulted in a fall in energy barrier from $\sim 12kT$ to $\sim 1kT$ for a pore radius increase of ~1 Å. Similar patterns were seen for the 1OED structure and for the other key rings of residues (i.e. 9′ and 17′).

*3.3. Opening the pore*

Having established that small changes in pore radius resulting from alanine substitution can have a substantial effect on the electrostatic profile, we decided to explore the effect of incremental increases in pore radius at the key sidechain rings (i.e. 9′, 13′, and 17′) on the energy profile. This was done by using an approach previously employed for modeling opening of the pore of the bacterial potassium channel KcsA (Biggin and Sansom, 2002), whereby incremental expansion of van der Waals particles within the pore constriction was used to drive the channel from a closed to a (more) open state. To this end, the radius of three van der Waals spheres placed within the pore of the 2BG9 TM region (see Methods for details) was increased from $\rho = 1$ to 7.2, with relaxation of the pore structure by energy minimization following each step. The pore radius profiles (figure 3A) of the resultant structures show a steady increase in minimum radius from ~2.4 Å (for the unperturbed 2BG9 structure) to ~3.5 Å (for the $\rho = 7.2$ perturbed structure). Note that in the latter case the pore minimum corresponds to the hydrophilic 2′ sidechain ring; the hydrophobic sidechain rings (9′, 13′, and 17′) are increased in radius to >4 Å at the end of the perturbation. If we compare the electrostatic free energy profile between the closed state of the pore (i.e. 2BG9) and a maximally opened state of the channel (e.g. the $\rho = 7.2$ perturbed structure; figure 3B) we see that the barriers due to the hydrophobic gates at 9′, 13′, and 17′ have disappeared, and even the barriers due to sidechain dipoles at 2′ and 6′ (see above) are greatly reduced.

Having perturbed the radius of the rings of hydrophobic gating sidechains (9′, 13′, and 17′) it is interesting to explore the relationship between the pore radius in the vicinity of each of these rings and the corresponding electrostatic barrier heights for the various incrementally perturbed structures (figure 4). For each of the three rings there is an approximately linear relationship between the barrier height and the pore radius (figure 4A), at least until the pore is fully open (i.e. the barrier height falls significantly below $2kT$).

Using a cutoff of $2kT$ to define a barrier to Na$^+$ ion permeation, we can see that an increase in pore radius of ~1.5 Å is needed in the vicinity of the 9′ and 13′ hydrophobic sidechain rings to open the pore (figure 4B). In contrast, the 17′ ring only needs to be expanded by ~0.4 Å to be opened. Thus we may conclude that the 9′ and 13′ rings form the main gates within the nAChR pore. The 9′ ring in particular has been implicated in gating via a number of mutational studies, e.g. (Bertrand *et al.*, 1993; Labarca *et al.*, 1995). We note that in addition to barriers observed in the 9′ to 17′ region, there is also a energy barrier at position 2′, which was not readily perturbed using the van der Waal spheres approach. However, we also note that this barrier is due to a ring of polar sidechains (serine and threonine) and so is likely to disappear as a result of sidechain conformational relaxation in the presence of a permeant ion (Sansom, 1992).

**4. Conclusion**

The clear implication from the results of this study is that a relatively small increase in the radius of the hydrophobic gate ($\Delta r \approx 1.5$ Å) is sufficient to open the pore of the nAChR (figure 5). This is consistent with earlier suggestions based on simulations of water in simple models of nanopores (Beckstein *et al.*, 2001; Beckstein and Sansom, 2003, 2004) and with recent experimental studies (Cymes *et al.*, 2005; Cymes and Grosman, 2008) which suggest that there is minimal rotation of the M2 helices upon opening of the channel, and that the widening of the pore in the plane of the membrane is not greater than a few Ångströms. Our results are also consistent with Brownian dynamics simulations on an open state model of the pore (Corry, 2006) which was constructed with a increase in pore radius of ~1.5 Å and which yielded an open state conductance consistent with that obtained experimentally. Recent MD simulations of a homology model of a human nAChR also suggest that an increase in channel radius of ~1 Å could result in channel opening (Wang *et al.*, 2008). Furthermore, the recent X-ray structure of a prokaryotic homologue of the nAChR (ELIC from *Erwinia chrysanthemi*) (Hilf and Dutzler, 2008) in a nonconductive conformation also suggests that a pore expansion, rather than M2 helix rotation or twist models (Taly *et al.*, 2005; Cheng *et al.*, 2006; Cheng *et al.*, 2007; Liu *et al.*, 2008), may underlie gating. The pore in ELIC is physically occluded. However, this may reflect a more stable closed-channel conformation than that in the *Torpedo* nAChR, because ELIC is missing the intracellular domain present in the vertebrate nAChR or because of the relative low sequence identity (<20%) between the nAChR and ELIC.

In addition to providing insights into to molecular basis of the function of the nAChR, these studies also have implications for the design of gated nanopores (Beckstein *et al.*, 2001; Peter and Hummer, 2005). Clearly a combined computational and experimental approach will be needed in order to design hydrophobic gates in novel or existing nanopores.

It is important to reflect critically upon the methodology used in these studies. The first aspect is the use of continuum electrostatics calculations to estimate the barrier height of the hydrophobic gate in the models of the nAChR pore. Although it is in some circumstances more accurate to estimate such barriers via molecular dynamics simulations to yield potentials of mean force (i.e. free energy profiles) (Amiri *et al.*, 2005; Beckstein and Sansom, 2006; Ivanov *et al.*, 2007) such studies are computationally demanding, and limit the number of model structures which may realistically be compared. Comparison of continuum electrostatics and MD potential of mean force (Beckstein *et al.*, 2004; Amiri *et al.*, 2005) suggests that although continuum (i.e. Poisson-Boltzmann) calculations may underestimate the barrier for radii greater than 4 Å, for the range applicable here (~2.5 to ~4 Å) the two approaches are in reasonable agreement. We have also not calculated van der Waals interactions between the ion and the gating residues. The contribution of such van der Waals interactions to the ion's potential energy profile were previously shown only to be of significance at pore radii of < ~2 Å (Tai *et al.*, 2008) and so omitting this term is unlikely to make any significant difference to the results.

In summary, these models and calculations provide further insight into possible gating mechanisms of the nAChR in particular, and of nanopores in general. They provide a testable hypothesis concerning the degree of pore expansion required to switch a hydrophobic gate.

**Acknowledgments**

We thank Nathan Baker for the APBS software, Philip Biggin for advice and helpful discussions, and the Wellcome Trust and the Biotechnology and Biological Sciences Research Council (United Kingdom) for funding.

**References**

Allen R, Hansen J P and Melchionna S 2003 Molecular dynamics investigation of water permeation through nanopores *J. Chem. Phys.* **119** 3905-19
Amiri S, Tai K, Beckstein O, Biggin P C and Sansom M S P 2005 The α7 nicotinic acetylcholine receptor: molecular modelling, electrostatics, and energetics. *Mol. Memb. Biol.* **22** 151-62
Anishkin A and Sukharev S 2004 Water dynamics and dewetting transitions in the small mechanosensitive channel MscS *Biophys. J.* **86** 2883-95
Archakov A I and Ivanov Y D 2007 Analytical nanobiotechnology for medicine diagnostics *Molec. Biosys.* **3** 336-42
Ashcroft F M 2000 *Ion Channels and Disease* (San Diego: Academic Press)
Baker N A, Sept D, Joseph S, Holst M J and McCammon J A 2001 Electrostatics of nanosystems: application to microtubules and the ribosome *Proc. Nat. Acad. Sci. USA* **98** 10037-41
Beckstein O, Biggin P C, Bond P J, Bright J N, Domene C, Grottesi A, Holyoake J and Sansom M S P 2003 Ion channel gating: insights via molecular simulations *FEBS Lett.* **555** 85-90
Beckstein O, Biggin P C and Sansom M S P 2001 A hydrophobic gating mechanism for nanopores *J. Phys. Chem. B* **105** 12902-5
Beckstein O and Sansom M S P 2003 Liquid–vapor oscillations of water in hydrophobic nanopores *Proc. Nat. Acad. Sci. USA* **100** 7063-8
Beckstein O and Sansom M S P 2004 The influence of geometry, surface character and flexibility on the permeation of ions and water through biological pores *Phys. Biol.* **1** 42-52
Beckstein O and Sansom M S P 2006 A hydrophobic gate in an ion channel: the closed state of the nicotinic acetylcholine receptor *Phys. Biol.* **3** 147-59
Beckstein O, Tai K and Sansom M S P 2004 Not ions alone: barriers to ion permeation in nanopores and channels *J. Amer. Chem. Soc.* **126** 14694-5
Bertrand D, Galzi J L, Devillers-Thiéry A, Bertrand S and Changeux J P 1993 Stratification of the channel domain in neurotransmitter receptors *Curr. Opin. Cell Biol.* **5** 688-93.

Biggin P C and Sansom M S P 2002 Open-state models of a potassium channel *Biophys. J.* **83** 1867-76
Bratko D, Daub C D, Leung K and Luzar A 2007 Effect of field direction on electrowetting in a nanopore *J. Amer. Chem. Soc.* **129** 2504-10
Chen M, Khalid S, Sansom M S P and Bayley H 2008 Outer membrane protein G: Engineering a quiet pore for biosensing *Proceedings of the National Academy of Sciences of the United States of America* **105** 6272-7
Cheng X, Ivanov I, Wang H, Sine S M and McCammon A 2007 Nanosecond-timescale conformational dynamics of the human α7 nicotinic acetylcholine receptor *Biophys. J.* **93** 2622-32
Cheng X, Lu B, Grant B, Law R J and McCammon J A 2006 Channel opening motion of α7 nicotinic acetylcholine receptor as suggested by normal mode analysis *J. Mol. Biol.* **355** 310-24
Corringer P J, Le Novere N and Changeux J P 2000 Nicotinic receptors at the amino acid level *Ann. Rev. Pharmacol. Toxicol.* **40** 431-58
Corry B 2006 An energy-efficient gating mechanism in the acetylcholine receptor channel suggested by molecular and Brownian dynamics *Biophys. J.* **90** 799-810
Corry B 2008 Designing carbon nanotube membranes for efficient water desalination *J. Phys. Chem. B* **112** 1427-34
Cruz-Chu E R, Aksimentiev A and Schulten K 2006 Water-silica force field for simulating nanodevices *J. Phys. Chem. B* **110** 21497-508
Cymes G D and Grosman C 2008 Pore-opening mechanism of the nicotinic acetylcholine receptor evinced by proton transfer *Nature Struct. Molec. Biol.* **15** 389-96
Cymes G D, Ni Y and Grosman C 2005 Probing ion-channel pores one proton at a time *Nature* **438** 975-80
Dolinsky T J, Nielsen J E, McCammon J A and Baker N A 2004 PDB2PQR: an automated pipeline for the setup, execution, and analysis of Poisson-Boltzmann electrostatics calculations *Nucl. Acids Res.* **32** W665-7
Fortina P, Kricka L J, Surrey S and Grodzinski P 2005 Nanobiotechnology: the promise and reality of new approaches to molecular recognition *Trends Biotech.* **23** 168-73
Hilf R J C and Dutzler R 2008 X-ray structure of a prokaryotic pentameric ligand-gated ion channel *Nature* **452** 375-9
Hille B 2001 *Ionic Channels of Excitable Membranes* (Sunderland, Mass.: Sinauer Associates Inc.)
Hummer G, Pratt L R and García A E 1997 Ion sizes and finite-size corrections for ionic-solvation free energies *J. Chem. Phys.* **107** 9275-7
Hummer G, Rasaiah J C and Noworyta J P 2001 Water conduction through the hydrophobic channel of a carbon nanotube *Nature* **414** 188-90
Humphrey W, Dalke A and Schulten K 1996 VMD - Visual Molecular Dynamics *J. Molec. Graph.* **14** 33-8
Ivanov I, Cheng X, Sine S M and McCammon J A 2007 Barriers to ion translocation in cationic and anionic receptors from the Cys-loop family *J. Amer. Chem. Soc.* **129** 8217-24
Labarca C, Nowak M W, Zhang H, Tang L, Deshpande P and Lester H A 1995 Channel gating governed symmetrically by conserved leucine residues in the M2 domain of nicotinic receptors *Nature* **376** 514-6

Lester H 1992 The permeation pathway of neurotransmitter-gated ion channels *Ann. Rev. Biophys. Biomol. Struct.* **21** 267-92.
Lester H A, Dibas M I, Dahan D S, Leite J F and Dougherty D A 2004 Cys-loop receptors: new twists and turns *Trends Neurosci.* **27** 329-36
Lindahl E, Hess B and van der Spoel D 2001 GROMACS 3.0: a package for molecular simulation and trajectory analysis *J. Molec. Model.* **7** 306-17
Liu X L, Xu Y C, Li H L, Wang X C, Jiang H L and Barrantes F J 2008 Mechanics of channel gating of the nicotinic acetylcholine receptor *PLoS Comput. Biol.* **4**
Lu D Y, Aksimentiev A, Shih A Y, Cruz-Chu E, Freddolino P L, Arkipov A and Schulten K 2006 The role of molecular modelling in bionanotechnology *Phys. Biol.* **3** S40-S53
Miyazawa A, Fujiyoshi Y and Unwin N 2003 Structure and gating mechanism of the acetylcholine receptor pore *Nature* **423** 949-55
Peter C and Hummer G 2005 Ion transport through membrane-spanning nanopores studied by molecular dynamics simulations and continuum electrostatics calculations *Biophys. J.* **89** 2222-34
Roux B and Schulten K 2004 Computational studies of membrane channels *Structure* **12** 1343-51
Sansom M S P 1992 The roles of serine and threonine sidechains in ion channels: a modelling study. *Eur. Biophys. J.* **21** 281-98
Scott W R P, Hunenberger P H, Tironi I G, Mark A E, Billeter S R, Fennen J, Torda A E, Huber T, Kruger P and van Gunsteren W F 1999 The GROMOS biomolecular simulation program package *J. Phys. Chem. A* **103** 3596-607
Smart O S, Neduvelil J G, Wang X, Wallace B A and Sansom M S P 1996 Hole: A program for the analysis of the pore dimensions of ion channel structural models. *J. Mol. Graph.* **14** 354-60
Tai K, Haider S, Grottesi A and Sansom M S P 2008 Ion channel gates: comparative analysis of energy barriers *Eur. Biophys. J.* 10.1007/s00249-008-0377-x
Takaiwa D, Hatano I, Koga K and Tanaka H 2008 Phase diagram of water in carbon nanotubes *Proc. Natl. Acad. Sci. USA* **105** 39-43
Taly A, Delarue M, Grutter T, Nilges M, Le Novere N, Corringer P J and Changeux J P 2005 Normal mode analysis suggests a quarternary twist model for the nicotinic receptor gating mechanism *Biophys. J.* **88** 3954-65
Unwin N 2005 Refined structure of the nicotinic acetylcholine receptor at 4 Å resolution *J. Mol. Biol.* **346** 967–89
Wang H L, Cheng X L, Taylor P, McCammon J A and Sine S M 2008 Control of cation permeation through the nicotinic receptor channel *PLoS Comput. Biol.* **4**
Wang J, Cieplak P and Kollman P A 2000 How well does a restrained electrostatic potential (RESP) model perform in calculating conformational energies of organic and biological molecules? *J. Comput. Chem.* **21** 1049-74

**Table 1.** Sequences of M2 helices. The sequences of the hydrophobic gate region of the M2 helix for each of the nAChR five chains is shown, where α, β, γ and δ are the four different subunit types. This provides an alignment of the sequences of the M2 residues that line the pore for residues 2′-17′ (*47*). The putative gate regions (9′, 13′, and 17′) formed by hydrophobic residues create a hydrophobic and narrow pore surface between the M2 helices.

| *Subunit* | \multicolumn{16}{c}{*M2 sequence*} | *Residues* |
|---|---|---|---|---|---|---|---|---|---|---|---|---|---|---|---|---|---|
|   | 2′ | 3′ | 4′ | 5′ | 6′ | 7′ | 8′ | **9′** | 10′ | 11′ | 12′ | **13′** | 14′ | 15′ | 16′ | **17′** |   |
| α | T | L | S | I | S | V | L | **L** | S | L | T | **V** | F | L | L | **V** | 251 to 259 |
| β | S | L | S | I | S | A | L | **L** | A | L | T | **V** | F | L | L | **L** | 257 to 265 |
| γ | T | L | S | I | S | V | L | **L** | A | Q | T | **I** | F | L | F | **L** | 259 to 267 |
| δ | S | T | A | I | C | V | L | **L** | A | Q | A | **V** | F | L | L | **L** | 265 to 273 |

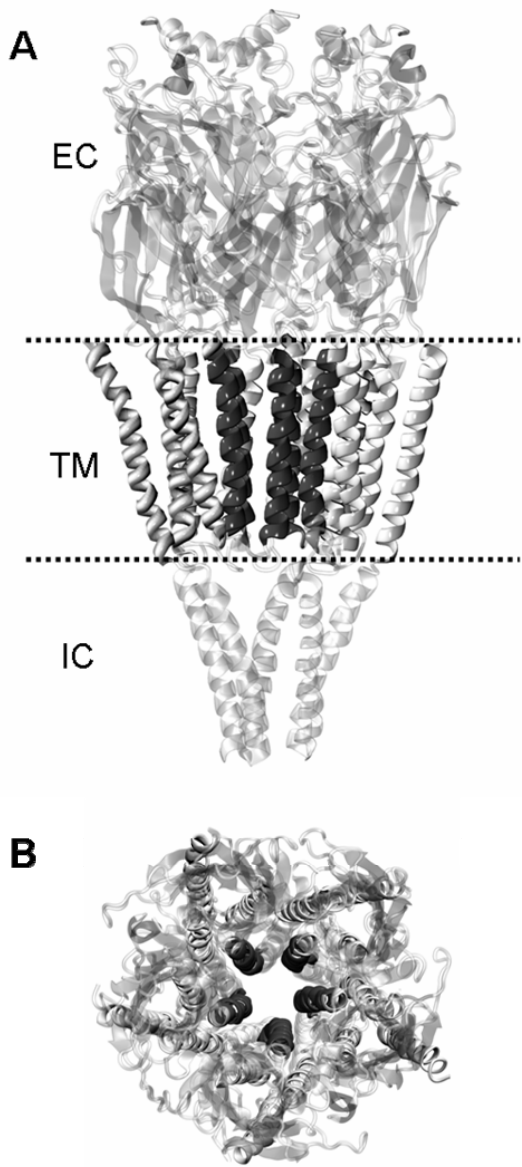

**Figure 1.** Diagram of the nAChR illustrating the location of the domains and the bilayer. **A** Diagram of the complete receptor (PDB entry 2BG9) showing the extracelllular (EC), transmembrane (TM), and intracellular (IC) domains. The approximate location of the lipid bilayer is indicated by the horizontal broken lines. **B** A view down the pore axis, with the pore-lining M2 helices shown in dark grey.

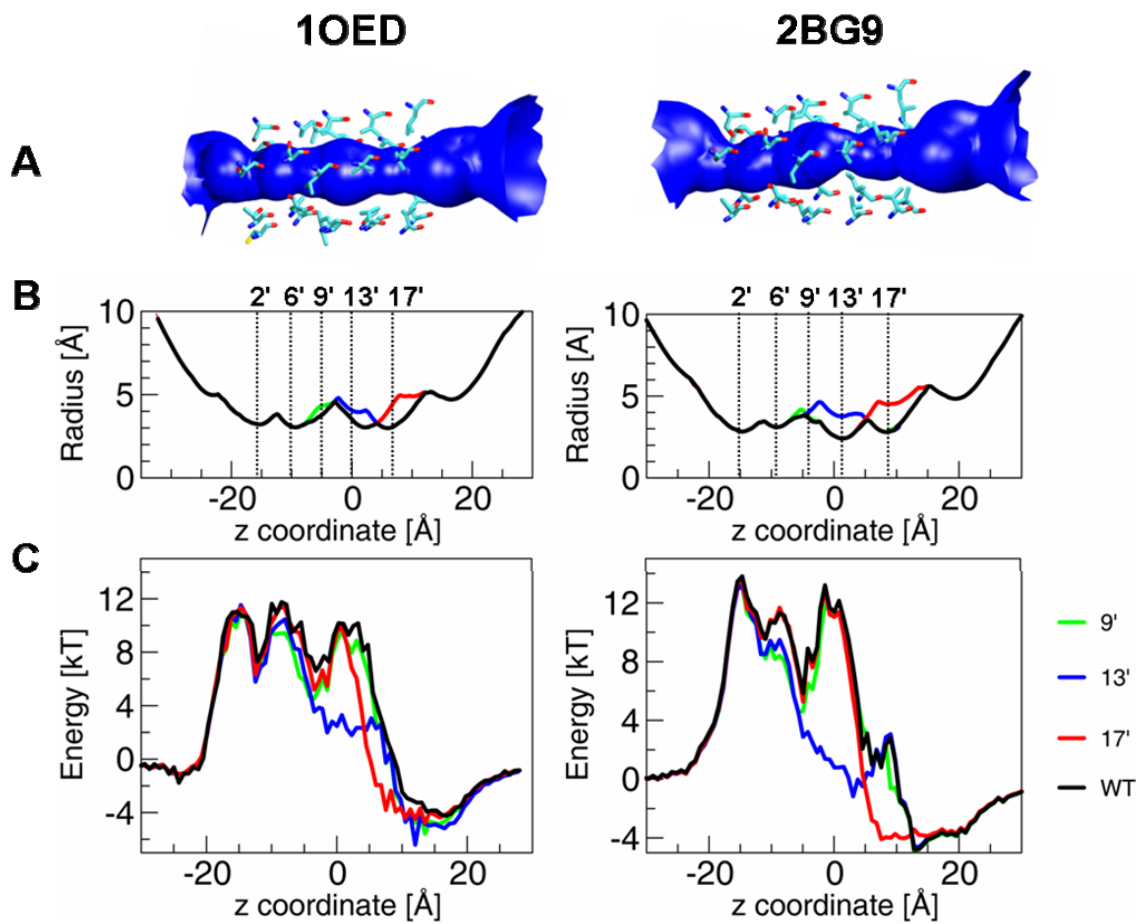

**Figure 2.** Pore radius and electrostatic energy (lower graphs) profiles for alanine substitution mutations of residues 9′, 13′, and 17′) of the M2 helices of the nAChR, for structures 1OED (left) and 2BG9 (right). The WT profiles (black lines) are provided for reference and correspond to those in figure 1. The other lines correspond to alanine substitution mutations of key hydrophobic residues (green = 9′, blue = 13′, and red = 17′). In each case a diagram of the sidechains of key residues from the M2 helices forming the pore (bonds format) and of the pore lining surface (blue) is shown (**A**) above graphs of the (**B**) pore radius and (**C**) electrostatic energy profiles. Note that the central hydrophobic sidechain ring (i.e. residues 13′) corresponds approximately to the centre ($z = 0$) of the coordinate system.

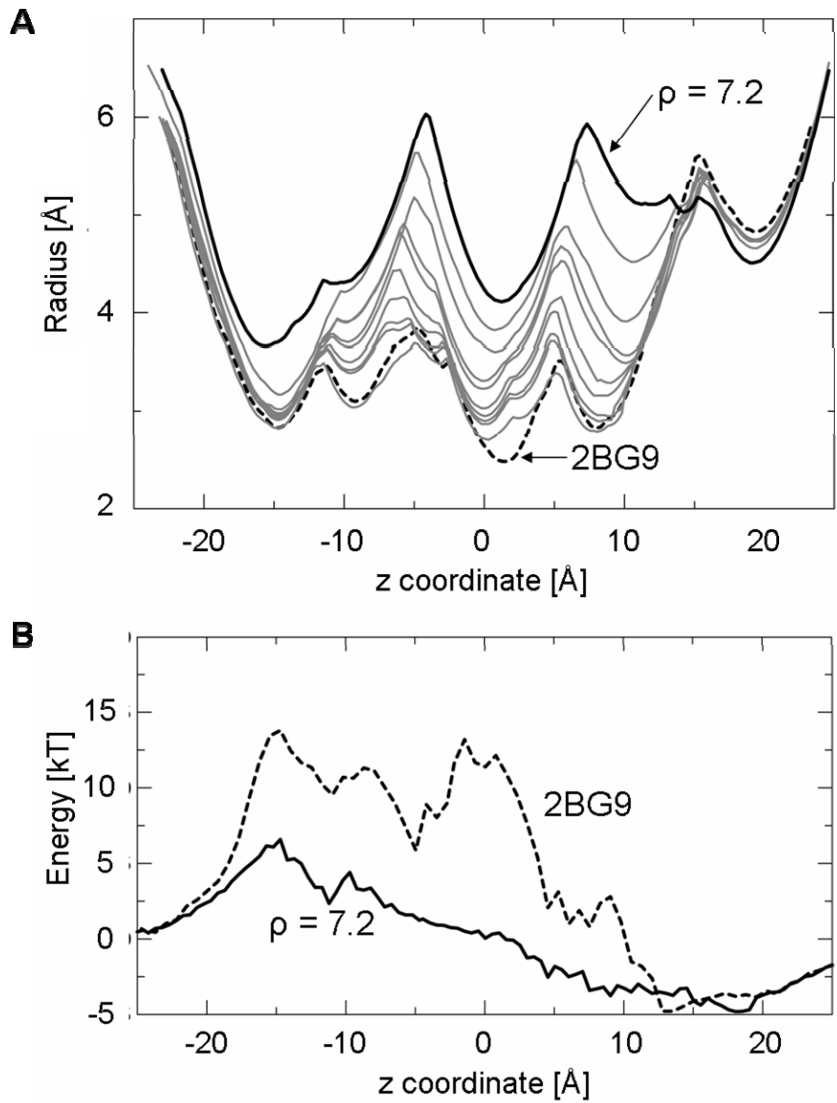

**Figure 3.** **A** Pore radius profiles for the nAChR pore expanded by van der Waals spheres of incrementally increasing radius (see Methods for details). The broken black line corresponds to the unperturbed 2BG9 pore; the solid black line corresponds to the maximally perturbed ($\rho = 7.2$) pore; and the solid grey lines correspond to the incrementally expanded pore models in between these two extremes. **B** Electrostatic energy profiles for unperturbed 2BG9 (broken black line) and the maximally perturbed ($\rho = 7.2$; solid black line) pores compared.

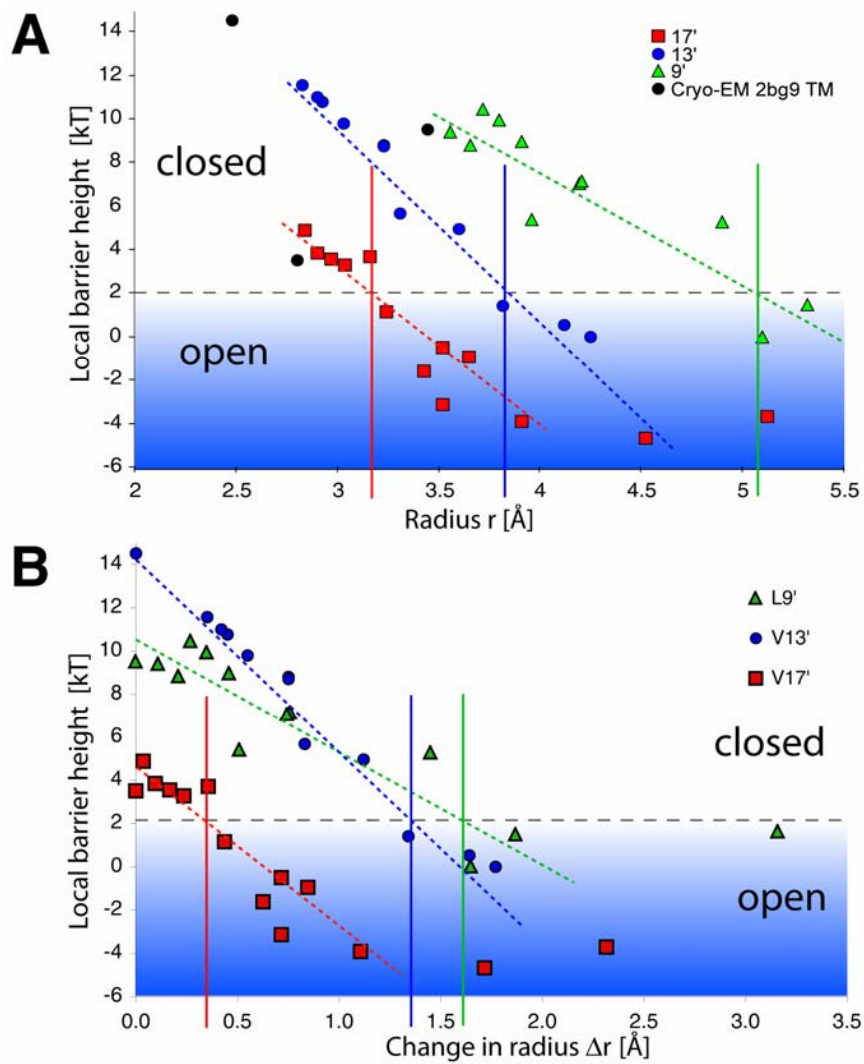

**Figure 4.** The relationship between the local pore/barrier radius and the local electrostatic energy barrier height for the 9′, 13′, and 17′ sidechain rings. **A** The electrostatic free energy of a $Na^+$ ion placed at positions 9′ (green), 13′ (blue), and 17′ (red) as a function of the pore radius for the models in which the 2BG9 pore was incrementally expanded (see figure 3 and text for details). **B** The relationship between the *change* in Δ*r* pore radius and the electrostatic energy of a $Na^+$ ion placed at positions 9′ (green), 13′ (blue), and 17′ (red).

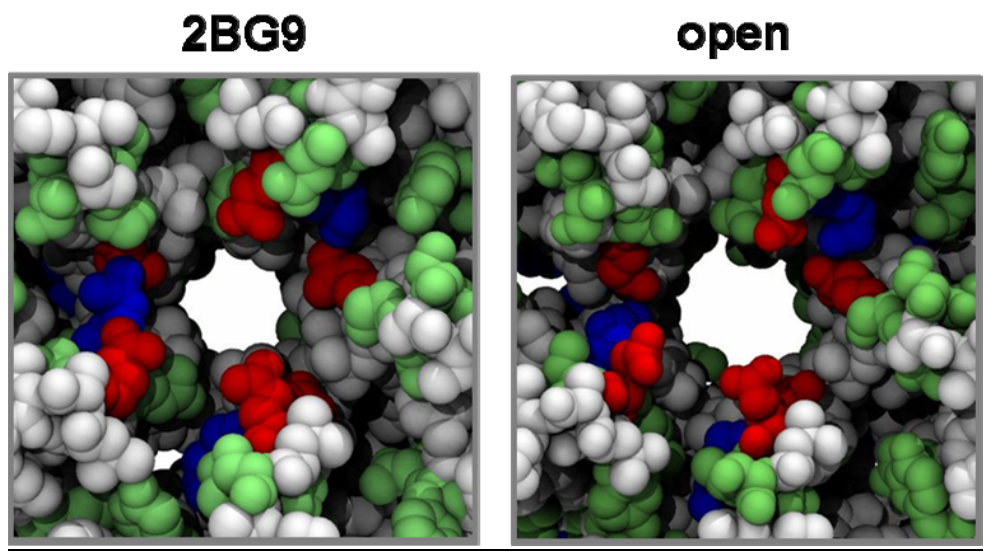

**Figure 5.** Structure of the gate region (viewed from the extracellular mouth of the pore) for the closed (2BG9) and open (perturbed; $\rho = 7.2$) models of the nAChR pore. Hydrophobic residues are rendered white, polar ones green, acidic sidechains in red, and basic ones in blue.